\documentclass{amsart}

\usepackage[english]{babel} 
\usepackage[latin1]{inputenc}
\usepackage[T1]{fontenc} 

\usepackage{verbatim}

\usepackage{graphicx}

\renewcommand{\det}[1]{{\frac{\partial {#1}}{\partial t}}}
\newcommand{\der}[1]{{\frac{\partial {#1}}{\partial r}}}
\newcommand{\derr}[1]{{\frac{\partial^2 {#1}}{\partial r^2}}}

\newcommand{\Pu}{p_c}
\newcommand{\Pd}{p_i}
\newcommand{\Pt}{p_i}

\newcommand{\Cu}{c_c}
\newcommand{\Cd}{c_i}
\newcommand{\Ct}{c_i}

\newcommand{\graffa}[2][lclll]{\left\{\begin{array}{#1} #2\\ \end{array} \right.}

\begin{document}

\title[A simple model of filtration and macromolecule transport]{A simple model of filtration and macromolecule transport through microvascular walls}

\author{Laura Facchini}
\address{Department of Mathematics, University of Trento (Italy) \\ via Sommarive 14, 38123 Trento (TN)}
\email{laura.facchini@unitn.it}

\author{Alberto Bellin}
\address{Department of Civil, Environmental and Mechanical Engineering, University of Trento (Italy) \\ via Mesiano 77, 38123 Trento (TN)}
\email{alberto.bellin@unitn.it}

\author{Eleuterio F. Toro}
\address{Laboratory of Applied Mathematics. DICAM, University of Trento (Italy) \\ via Mesiano 77, 38123 Trento (TN)}
\email{eleuteriofrancisco.toro@unitn.it}

%\thanks{This work is partially funded by CARITRO (\emph{Fondazione Cassa di Risparmio di Trento e Rovereto}, Italy).}

\subjclass{ % http://www.ams.org/mathscinet/msc/msc2010.html
82C70; % Transport processes
65C20 % Models, numerical methods
}
 
\keywords{Ultrafiltration; Starling's law; Capillary wall; Nonlinear transport of macromolecules}

\begin{abstract}
Multiple Sclerosis (MS) is a disorder that usually appears in adults in their thirties. It has a prevalence that ranges between 2 and 150 per 100 000. Epidemiological studies of MS have provided hints on possible causes for the disease ranging from genetic, environmental and infectious factors to other factors of vascular origin. Despite the tremendous effort spent in the last few years, none of the hypotheses formulated so far has gained wide acceptance and the causes of the disease remain unknown. From a clinical point of view, a high correlation has been recently observed between MS and Chronic Cerebro-Spinal Venous Insufficiency (CCSVI) in a statistically significant number of patients. In this pathological situation CCSVI may induce alterations of blood pressure in brain microvessels, thereby perturbing the exchange of small hydrophilic molecules between the blood and the external cells. In the presence of large pressure alterations it cannot be excluded also the leakage of macromolecules that otherwise would not cross the vessel wall. All these disorders may trigger immune defenses with the destruction of myelin as a side effect.  In the present work we investigate the role of perturbed blood pressure in brain microvessels as  driving force for an altered exchange of small hydrophilic solutes and  leakage of macromolecules into the interstitial fluid. With a simplified, yet realistic, model we obtain closed-form steady-state solutions for fluid flow and solute transport across the microvessel wall. Finally, we use these results (i) to interpret experimental data available in the literature and (ii) to carry out a preliminary analysis of the disorder in the exchange processes  triggered by  an increase of blood pressure, thereby relating our preliminary results to the hypothesised vascular connection to MS.
\end{abstract}

\maketitle

\section{INTRODUCTION}

Multiple Sclerosis is an autoimmune neurodegenerative disorder of unknown origin that damages the myelin, a fatty layer that envelops and protects the axons. The damage of the myelin causes  neuron electrical impulses to travel slowly along their axons, leading to a variety of symptoms with debilitating consequences. The repeated damage of the myelin causes the loss of the remyelination capacity of oligodendrocytes and produces scar-like lesions around damaged axons. From a clinical point of view, these lesions are demonstrated to be localised in the white matter and  to be venocentric (i.e. these plaques are always found around venules). 

Recent clinical evidence suggests an association of MS with CCSVI [Zamboni, 2006; Singh and Zamboni, 2009; Zamboni et al., 2009]. However, the evolution of the process from venous stenosis to local hypertension and leakage of hematic substance, which may trigger the immune response as ultimate cause of demyelination and neurodegeneration typical of MS has been the subject of an intense debate with often opposed views. This is a controversial hypothesis, yet it is consistent with the predominantly venocentric orientation of the MS inflammatory lesions and with the otherwise unexplained perivenular iron deposition observed in many clinical cases [Adams, 1988]. A possible path connecting cerebrospinal venous stenosis to chronic fatigue and MS was proposed by Tucker [2011] on the basis of qualitative considerations of elementary fluid mechanics. 

From a physiological point of view, the microvessel wall plays an important role in maintaining the equilibrium between intravascular and extravascular fluid compartments. Under normal conditions the vessel walls are nearly impermeable to macromolecules, while lipophilic species and small hydrophilic substances are allowed to cross the wall and reach the surrounding tissue. Fluid flow and transport of dissolved molecules across the walls depend on the permeability and diffusivity of the membrane composing the wall. Therefore, alterations of the blood pressure may lead to impaired exchange processes and, in extreme cases, to leakage of hematic fluid. Several alterations of these exchange processes have been observed, mainly in compartments other than the brain, resulting in leakage of macromolecules, which is typically attributed to reduction of osmotic pressure, or inflammatory processes that alter the endothelial structure. 

In the present work we investigate the role of an altered blood pressure as the driving force for alteration of the exchange processes and the  leakage of macromolecules. In particular, we analyze through a simplified, yet realistic, flow and transport model, the impact of alterations in the hydrostatic blood pressure on transport of molecules across the microvessel wall. The microvessel wall is assumed to be composed of two layers with different permeability and porosity, as assumed in previous studies on fluid flow and macromolecules transport in heteroporous membranes.  The inner layer represents the glycocalyx, a membrane composed of extracellular polymeric material which is believed to exert an important sieving effect on macromolecules, while the external layer represents the combined effect of the endothelial cells, the basal membrane and the external astrocyte feet.     

With this model we obtain closed-form steady-state solutions for the fluid flow and solute transport through the microvessel walls, which can be used for a preliminary analysis of the leakage of macromolecules due to an increase of blood pressure in CCSVI/MS patients. 

\section{MICROVESSEL ANATOMY}

In this section we summarise anatomic features of mammalian blood vessels useful to describe the geometry of the computational domain used in the present work. 

The circulatory system is composed by vessels of size ranging from centimetres in the main ones to a few microns in the capillary bed. The structure of the vessel wall differs between arteries and veins and also between large vessels and capillaries. Each segment of circulation shows an optimal combination of size, wall composition, thickness and cross-sectional area that best fulfils its function. For example, arteries are more muscular than veins because they have to bear the pumping force of the heart. 

Large vessels are formed of three layers: the endothelium, the middle layer composed by smooth muscle cells and the connective layer.
On the contrary, small vessels such as capillaries, venules and arterioles are only one-cell thick, in order to optimize the exchange of small hydrophilic molecules from the blood stream to the interstitial volume before crossing the cell membrane.

Molecules dissolved in water are driven through the vessel wall by the gradient of the net pressure $p$, which is given by the difference between the hydrostatic $P$ and osmotic $\Pi$ pressure: $p=P- \sigma \Pi$, where $\sigma$ is the reflection coefficient. $\sigma$ depends on the ratio between the Stokes radius of the molecule and the pore radius, or the size of the cleft between adjacent endothelial cells. When the size of the molecule is comparable with the pore size (or the aperture), the vessel wall behaves as a perfect membrane and $\sigma \rightarrow 1$. On the other hand, when the molecules are much smaller than the pore size, the membrane effect vanishes and $\sigma \rightarrow 0$. In the latter case, transport across the vessel wall is controlled by the gradient of the hydrostatic pressure. For a given pore (or cleft) size, the role of the osmotic pressure increases with the size of molecule. In the present work we consider two layers, one represented by the glycocalyx and the other by the cleft. The sieving effect of glycocalyx on macromolecules is represented by a $\sigma$ value that approaches 1, while in the stratum representing the endothelial cells $\sigma$ is typically smaller, to reflect the larger aperture of the tight junctions connecting the two sides of the cleft at the border between adjacent cells [Levick, 2010]. 

\section{CONCEPTUAL MODEL}

Let us approximate the microvessel geometry as a rigid circular cylinder, infinitely long in the $z$-direction, i.e. in the direction of the blood stream. We assume the vessel wall composed by one or more permeable layers of a given thickness. Physical properties, such as permeability and molecular diffusion are assumed constant within a layer, but may vary across the layers. The porosity is assumed the same in all layers.
Molecules of a given Stokes radius are dissolved into the blood plasma at a concentration that does not modify its density and viscosity. Furthermore, to simplify the analysis we assume that the pressure gradient is small in the longitudinal direction, such that blood flow through the vessel lumen can be decoupled from the filtration through its wall. In general, the osmotic pressure changes with the solute concentration $c$. For small concentrations, the following linear relationship is often considered: $\Pi = \phi R T c$, where $\phi$ is a parameter that depends on the Stokes radius of the molecule, $R$ is the gas constant and $T$ is the absolute temperature. 
Consequently, the flow and transport equations are coupled through the concentration $c$ that feeds back through $\Pi$ to the flow. This leads Levick and Michel [2010] to conclude that microvessels cannot absorb fluid from the interstitial space, as is often argued. However, this feedback is important mainly when hydrostatic pressure is abruptly reduced, as in the Landis experiment \cite{Landis}, whereas here we are interested in the increase of hydrostatic pressure. We therefore neglect this feedback and solve the flow and transport equations separately.  

Under the above assumptions, mass balance of the solvent and the solute leads to the following governing equations for the pressure $p=p(x,y,z,t)$ 
\begin{equation} 
\det{p} = \frac{k \rho g}{\mu S_s} \nabla^2 p, 
\end{equation}
and for the concentration $c=c(x,y,z,t)$
\begin{equation} 
\det{c} + \frac{\mathbf{q}}{n} \cdot \nabla c = \nabla \cdot ( \mathbf{D} \cdot \nabla c ).
\end{equation}
where $k$ is the wall permeability, $\rho$ is the blood density, $g$ is the acceleration due to gravity, $\mu$ is the blood dynamic viscosity, $S_s$ is the specific storage of the porous material, $n$ is the porosity of the material and $\mathbf{D}$ is the diffusion tensor.

The specific water (solvent) discharge $\mathbf{q}=\mathbf{q}(x,y,z,t)$ is proportional to the net pressure gradient through the Starling equation \cite{levick:03}
\begin{equation} 
\mathbf{q} = -\frac{K}{\rho g} \nabla p, 
\end{equation} 
where 
\begin{equation} 
K = \frac{k \rho g}{\mu}.
\end{equation}  
Finally, the mass flux of the solute $\mathbf{f}_m=\mathbf{f}_m(x,y,z,t)$ is given by
\begin{equation} 
\mathbf{f}_m = (1-\sigma) \mathbf{q} c - n \mathbf{D} \cdot \nabla c.
\end{equation} 
The above equations written in cylindrical coordinates $(r,\theta,z)$ and assuming radial symmetry take the following form
\begin{eqnarray}
\det{p} &=& \frac{k \rho g}{\mu S_s} {1 \over r} {\partial \over \partial r} \left( r {\partial p \over \partial r} \right), \\
q       &=& -\frac{K}{\rho g} \der{p}, \\
\det{c} + \frac{q}{n} \der{c} &=& \left( {d \over r} + \der{d} \right) \der{c} + d \derr{c}, \\
f_m                           &=& (1-\sigma) q c - n d \der{c}.
\end{eqnarray}

Since the initial and boundary conditions are independent from the coordinates $z$ and $\theta$, we only consider the radial component $d$ of the diffusion tensor $\mathbf{D}$. 

Furthermore, we assume that $d$ is given by the sum of the molecular diffusion $d_m$ and the hydrodynamic dispersion $d_h = A q$, where $A$ is the dispersivity and $q$ is the radial component of the specific discharge $\mathbf{q}$.

In the next section we consider the steady-state solution of the above flow and transport equations.

\section{ANALYSIS}

The steady-state equations for the solvent and for the solute in cylindrical coordinates assume the following form
\begin{eqnarray}
0 &=& {\partial \over \partial r} \left( r {\partial p \over \partial r} \right), \\
\frac{q}{n} \der{c} &=& \left( {d \over r} + \der{d} \right) \der{c} + d \derr{c}.
\end{eqnarray}

\subsection{Steady-state solutions for a single-layer vessel wall}

\begin{figure}[ht]
\centering
$$\qquad \mbox{(a)} \qquad \qquad \qquad \qquad \qquad \mbox{(b)} \qquad $$
\includegraphics[width=4cm]{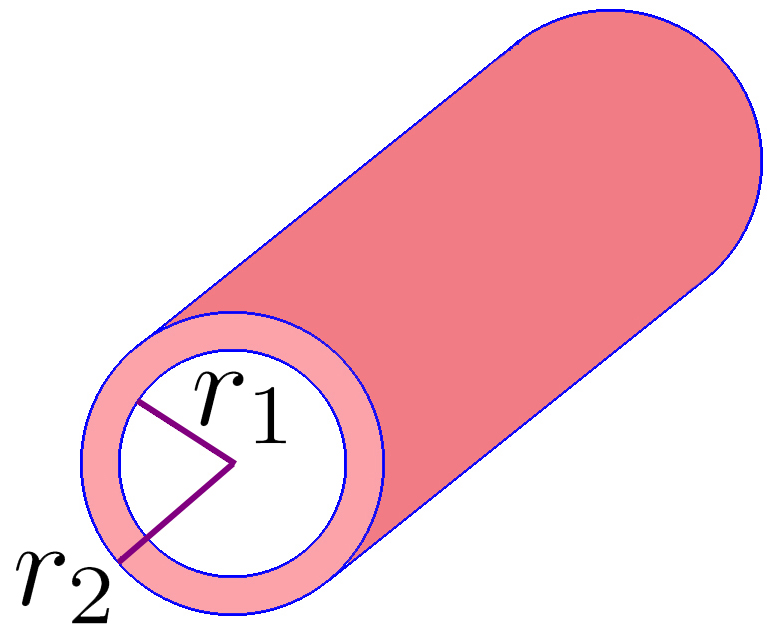} \qquad \includegraphics[width=3cm]{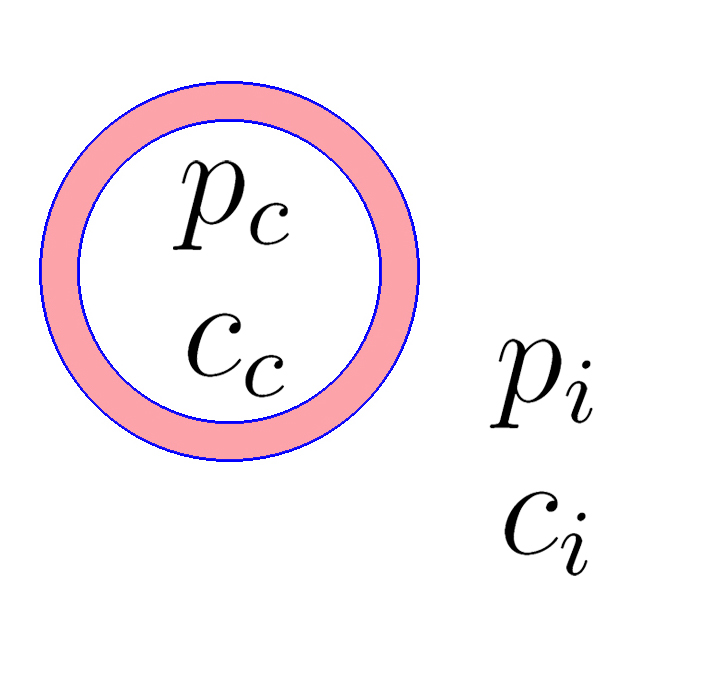} 
\smallskip
\caption{(a) The domain is an infinitely long hollow cylinder composed by one layer only, whose inner and outer radii are $r_1$ and $r_2$. 
(b) Cross section depicting boundary conditions, where $p_c = P_c - \sigma \Pi_c$ and $c_c$ refer to the net blood pressure and solute blood concentration, respectively and $p_i = P_i - \sigma \Pi_i$ and $c_i$ indicate the blood pressure and solute concentration in the interstitial fluid.}
\label{fig3}
\end{figure}

We now consider a geometrical situation as depicted in Figure \ref{fig3}(a), which shows a cylinder whose inner surface of the endothelial cells is represented by radius $r_1$ and whose outer wall is determined by radius $r_2$. 
 
In this case, we obtain two generic solutions
\begin{eqnarray}
p(r) &=& \alpha + \beta \ln r, \qquad r \in [r_1, r_2] \\
c(r) &=& \delta + \gamma h(r), \qquad r \in [r_1, r_2]
\end{eqnarray}
each of them depending on two parameters which can be computed by imposing the boundary conditions,
where $h(r)$ is an auxiliary function depending on the permeability and on the boundary conditions 
\begin{equation}
h(r)= \graffa{
{\displaystyle - \frac{\mu n}{k \beta} \left[ {\mu d_{m} r - k \beta A} \right]^{\displaystyle -\frac{k \beta}{n \mu d_{m}}}, } & \textrm{if } \quad k (\Pu - \Pd) \neq 0 \\ \\
\displaystyle \frac{\ln \left( \mu d_{m} r \right)}{d_{m}}, & \textrm{if } \quad k (\Pu - \Pd) = 0 } 
\end{equation}
for $r \in [r_1, r_2]$, with 
\begin{equation}
\beta = \frac{\Pu - \Pd}{\ln r_1 - \ln r_2}.
\end{equation}

We suppose that the boundary conditions are independent from the coordinates $z$ and $\theta$. So we set constant pressures and concentrations at the boundary, as depicted in Figure \ref{fig3}(b),
\begin{eqnarray}
p (r_1) &=& \Pu, \\ p (r_2) &=& \Pd, \\
c (r_1) &=& \Cu, \\ c (r_2) &=& \Cd,
\end{eqnarray}
where $p_c = P_c - \sigma \Pi_c$ refers to the net blood pressure, $c_c$ to the solute blood concentration, $p_i = P_i - \sigma \Pi_i$ indicates the blood pressure in the interstitial fluid and $c_i$ the interstitial solute concentration.
So we obtain closed-form steady-state solutions for $r \in [r_1, r_2]$, given by
\begin{eqnarray}
p(r) &=& \frac{\Pu \ln (r_2/r) + \Pd \ln (r/r_1)}{\ln (r_2 / r_1)}, \\
q(r) &=& - \frac{k}{\mu r} \frac{\Pu - \Pd}{\ln r_1 - \ln r_2}, \\ % NOW THE SIGN IS CORRECT
c(r) &=& \frac{\Cu [h(r)-h(r_2)] + \Cd [h(r_1)-h(r)]}{h(r_1) - h(r_2)}. \qquad
\end{eqnarray}

The mass flux depends on the values of the permeability and on the boundary conditions 
\begin{eqnarray}
\qquad f_m (r) &=& \graffa{
%\displaystyle - \frac{k \beta}{\mu r} \frac{\Cd h(r_1) - \Cu h(r_2)}{h(r_1) - h(r_2)}, & \textrm{ if } \quad k (\Cu - \Cd) \neq 0 WRONG \\ \\
\displaystyle \frac{k \beta}{\mu r} \left[ \sigma c(r) - \frac{\Cd h(r_1) - \Cu h(r_2)}{h(r_1) - h(r_2)} \right], & \textrm{ if } \quad k (\Cu - \Cd) \neq 0 \\ \\
\displaystyle - \frac{n}{r} \frac{\Cu - \Cd}{h(r_1) - h(r_2)},                         & \textrm{ if } \quad k (\Cu - \Cd) = 0 } 
\end{eqnarray}
for $r \in [r_1, r_2]$, with 
\begin{equation}
\beta = \frac{\Pu - \Pd}{\ln r_1 - \ln r_2}.
\end{equation}

\subsection{Steady-state solutions for a vessel wall composed by two layers}

\begin{figure}[ht]
\centering
$$\qquad \mbox{(a)} \qquad \qquad \qquad \qquad \qquad \mbox{(b)} \qquad $$
\includegraphics[width=3cm]{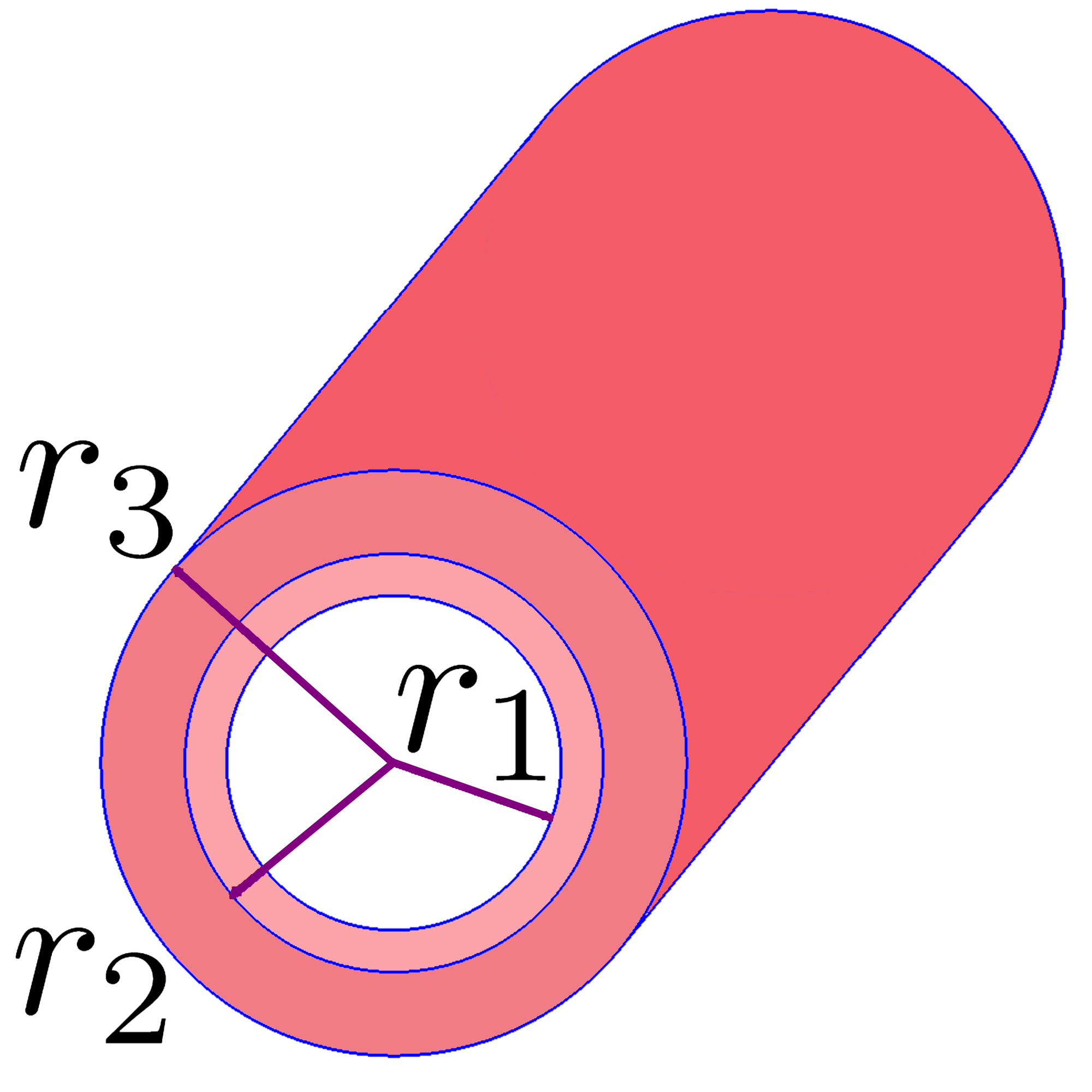} \qquad \includegraphics[width=4cm]{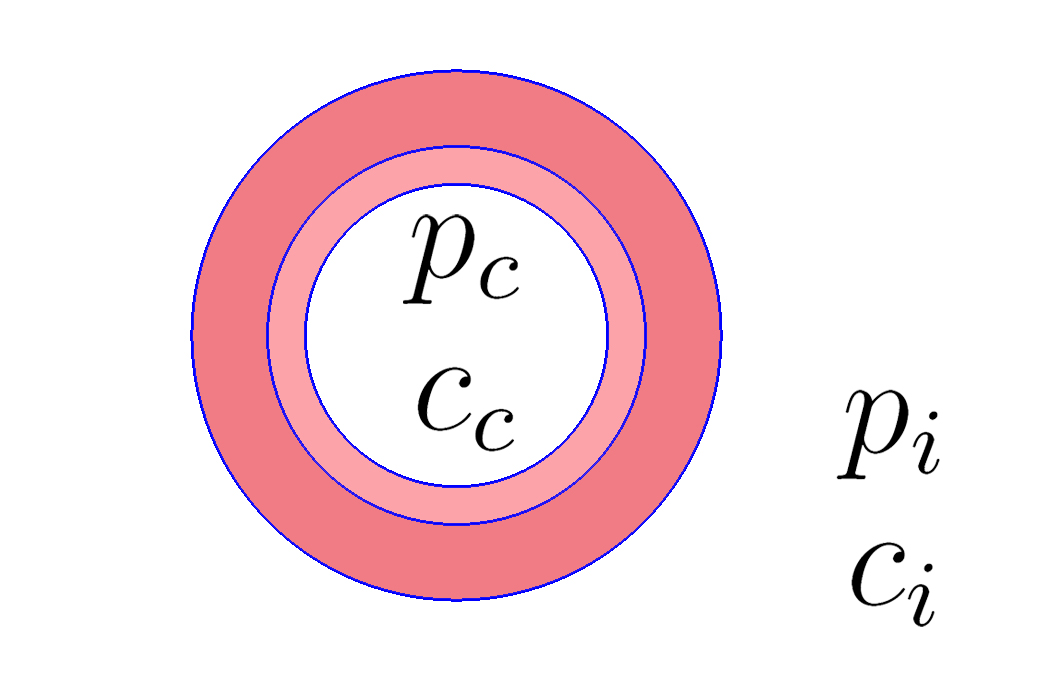} 
\smallskip
\caption{(a) The domain is an infinitely long hollow cylinder composed by two layers $(r_1, r_2)$ and $(r_2, r_3)$. 
(b) Cross section depicting boundary conditions, where $p_c = P_c - \sigma_1 \Pi_c$ and $c_c$ refer to the net blood pressure and solute blood concentration, respectively and $p_i = P_i - \sigma_2 \Pi_i$ and $c_i$ indicate the blood pressure and solute concentration in the interstitial fluid.}
\label{fig4}
\end{figure}

We now consider a more complex case in which the vessel wall is composed by two layers $(r_1, r_2)$ and $(r_2, r_3)$, as depicted in Figure \ref{fig4}(a), with different values of permeability, diffusivity and reflection coefficient. 

The general case of $m$ layers can be treated similarly. 

The generic solution for the net pressure assumes the following form 
\begin{equation}
\graffa{
p_1(r) = \alpha_1 + \beta_1 \ln r, & r \in [r_1, r_2] \\ 
p_2(r) = \alpha_2 + \beta_2 \ln r, & r \in [r_2, r_3] } 
\end{equation}
while the solute concentration is given by 
\begin{equation}
\graffa{
c_1(r) = \delta_1 + \gamma_1 h_1(r), & r \in [r_1, r_2] \\ 
c_2(r) = \delta_2 + \gamma_2 h_2(r), & r \in [r_2, r_3] } 
\end{equation}
where, similarly to the previous case, $h_j(r)$ is a function that depends on the geometry and the permeability of the layer 
\begin{eqnarray}
\qquad h_j(r) &=& \graffa{
\displaystyle - \frac{\mu n}{k_{eq} B} \left[ {\mu d_{m_j} r - k_{eq} B A_j} \right]^{\displaystyle -\frac{k_{eq} B}{n \mu d_{m_j}}}, & \textrm{if } \ k_1 k_2 (\Pu - \Pt) \neq 0 \\ \\
\displaystyle \frac{\ln \left( \mu d_{m_j} r \right)}{d_{m_j}}, & \textrm{if } \ k_1 k_2 (\Pu - \Pt) = 0 } 
\end{eqnarray}
for $r \in [r_j, r_{j+1}]$, recalling that $d_{m_j}$ is the molecular diffusion and $A_j$ is the dispersivity of the $j$-th layer. 
$B$ and $k_{eq}$ are now defined as
\begin{eqnarray}
B &=& \Pu - \Pt, \\
k_{eq} &=& \frac{k_1 k_2}{k_1 (\ln r_2 - \ln r_3) + k_2 (\ln r_1 - \ln r_2)}. \qquad
\end{eqnarray}

The constants appearing in the above solutions are obtained by imposing suitable boundary conditions for both the net pressure and solute concentration, as depicted in Figure \ref{fig4}(b)
\begin{eqnarray}
p_1 (r_1) &=& \Pu, \\ p_2 (r_3) &=& \Pt, \\
c_1 (r_1) &=& \Cu, \\ c_2 (r_3) &=& \Ct,
\end{eqnarray}
where $p_c = P_c - \sigma_1 \Pi_c$ refers to the net blood pressure, $c_c$ to the solute blood concentration, $p_i = P_i - \sigma_2 \Pi_i$ indicates the blood pressure in the interstitial fluid and $c_i$ the interstitial solute concentration.
These boundary conditions should be supplemented by the conditions resulting from imposing the continuity of the specific discharge and the solute flux at the interface between the two layers at $r=r_2$
\begin{eqnarray}
q_1 (r_2)     &=& q_2 (r_2), \\
f_{m,1} (r_2) &=& f_{m,2} (r_2),
\end{eqnarray}
and that both pressure and solute concentration are continuous at $r=r_2$
\begin{eqnarray}
p_1 (r_2) &=& p_2 (r_2), \\
c_1 (r_2) &=& c_2 (r_2).
\end{eqnarray}

With all these conditions the pressure within the first and second layer are given by
\begin{equation}
p_1(r) = \frac{k_2 \left[\Pt \ln (r_1 / r) - \Pu \ln (r_2 / r) \right] + \Pu k_1 \ln (r_2 / r_3)}{k_1 \ln (r_2 / r_3) + k_2 \ln (r_1 / r_2)}
\end{equation}
and
\begin{equation}
p_2(r) = \frac{k_1 \left[\Pt \ln (r_2 / r) - \Pu \ln (r_3 / r) \right] - \Pt k_2 \ln (r_2 / r_1)}{k_1 \ln (r_2 / r_3) + k_2 \ln (r_1 / r_2)}
\end{equation}
respectively.

The resulting expression for the specific discharge is the same in the two regions, indeed
\begin{eqnarray}
q(r) &=& - \frac{k_j}{\mu} \cdot \der{p_j}(r) \ = \ - \frac{k_1 k_2 (\Pu - \Pt)}{\mu[k_1 \ln (r_2 / r_3) + k_2 \ln (r_1 / r_2)]} \frac{1}{r}, \qquad 
\end{eqnarray}
for $r \in [r_1, r_3]$, where $j \in \{1,2\}$ indicates the layer we are considering.

Similarly, under steady-state conditions, the solute concentration assumes the following expression
\begin{equation}
c(r) = \graffa{
c_1(r) = \displaystyle \frac{S_1 + T_1 h_1(r)}{V}, & r \in [r_1, r_2] \\ \\
c_2(r) = \displaystyle \frac{S_2 + T_2 h_2(r)}{V}, & r \in [r_2, r_3] } 
\end{equation}
where the parameters $S_1, \ T_1, \ S_2, \ T_2, \ V$ depend on the value of $k_1 k_2 (\Pu - \Pt)$. Indeed, these parameters are defined as

\begin{equation}
S_1 = \graffa{
\Cu (1 + \sigma_1 - \sigma_2 ) h_1(r_2) h_2(r_3) - \Ct h_1(r_1) h_2(r_2) + \\
+ \Cu ( \sigma_2 - \sigma_1 ) h_1(r_2) h_2(r_2),     & \textrm{ if } \quad k_1 k_2 (\Pu - \Pt) \neq 0 \\ \\
-\Ct h_1(r_1) - \Cu [h_1(r_2) - h_2(r_2) + h_2(r_3)], & \textrm{ if } \quad k_1 k_2 (\Pu - \Pt) = 0 } 
\end{equation}

\begin{equation}
T_1 = \graffa{
\left[ \Ct - \Cu ( 1 - \sigma_1 + \sigma_2 ) \right] h_2(r_2) 
+ \Cu ( \sigma_2 - \sigma_1 ) h_2(r_3),              & \textrm{ if } \quad k_1 k_2 (\Pu - \Pt) \neq 0 \\ \\
\Cu - \Ct,                                           & \textrm{ if } \quad k_1 k_2 (\Pu - \Pt) = 0 } 
\end{equation}

\begin{equation}
S_2 = \graffa{
\Cu h_1(r_2) h_2(r_3) + \Ct ( \sigma_2 - \sigma_1 ) h_1(r_2) h_2(r_2) + \\
- \Ct (1-\sigma_1+\sigma_2) h_1(r_1) h_2(r_2),       & \textrm{ if } \quad k_1 k_2 (\Pu - \Pt) \neq 0 \\ \\
\Ct [h_1(r_1) - h_1(r_2) + h_2(r_2)] - \Cu h_2(r_3), & \textrm{ if } \quad k_1 k_2 (\Pu - \Pt) = 0 } 
\end{equation}

\begin{equation}
T_2 = \graffa{
\left[ \Ct (1 + \sigma_1 - \sigma_2 ) - \Cu \right] h_1(r_2) 
+ \Ct (\sigma_2-\sigma_1) h_1(r_1),                  & \textrm{ if } \quad k_1 k_2 (\Pu - \Pt) \neq 0 \\ \\
\Cu - \Ct,                                           & \textrm{ if } \quad k_1 k_2 (\Pu - \Pt) = 0 } 
\end{equation}

\begin{equation}
V = \graffa{
(\sigma_2-\sigma_1) [h_1(r_2) h_2(r_2) - h_1(r_1) h_2(r_3)] + \\
- (1-\sigma_1+\sigma_2) h_1(r_1) h_2(r_2) + \\
+ (1+\sigma_1-\sigma_2) h_1(r_2) h_2(r_3),           & \textrm{ if } \quad k_1 k_2 (\Pu - \Pt) \neq 0 \\ \\
h_1(r_1) - h_1(r_2) + h_2(r_2) - h_2(r_3),           & \textrm{ if } \quad k_1 k_2 (\Pu - \Pt) = 0. } 
\end{equation}

The resulting solute flux is the following
\begin{eqnarray}
\qquad f_{m,j}(r) &=& \graffa{
% \displaystyle \frac{k_{eq} B}{\mu r} \left[\sigma_j \frac{T_j}{V} h_j(r) - (1-\sigma_j) \frac{S_j}{V} \right], 
\displaystyle \frac{k_{eq} B}{\mu r} \left[\sigma_j c_j (r) - \frac{S_j}{V} \right],  & \textrm{ if } \quad k_1 k_2 (\Pu - \Pt) \neq 0 \\ \\
\displaystyle -\frac{n T_j}{V} \frac{1}{r},                                         & \textrm{ if } \quad k_1 k_2 (\Pu - \Pt) = 0 } 
\end{eqnarray}
for $r \in [r_j, r_{j+1}]$, where $j \in \{1,2\}$ indicates the layer we are considering and $\sigma_j$ represents the reflection coefficient that may be different in the two layers.

Similar expressions may be obtained for three and more layers.

\subsection{The travel time through the vessel wall}

An important quantity in the exchange process is the time a single solute molecule takes to cross the vessel wall. We call this time the travel time $\tau$, in analogy with transport in porous media. 

For the single layer case, $\tau$ may be approximated by neglecting the diffusive component of the mass flux 
\begin{eqnarray}
\tau &=& \int_{r_1}^{r_2} \frac{n}{(1-\sigma) q(r)} \ \textrm{d} \, r \ = \ %-\ \frac{n \mu}{\Pu - \Pd} \frac{\ln r_1 - \ln r_2}{k (1-\sigma)} \frac{r_2^2 - r_1^2}{2}. \\
-\ \frac{n \mu}{\beta k (1-\sigma)} \frac{r_2^2 - r_1^2}{2}.
\end{eqnarray}

\section{PRELIMINARY RESULTS} 

The structure of the vessels is very specialized in relation to their functionality and this specialization results in different permeability and reflection coefficients of the vessel wall. Table \ref{tab2} shows typical values of the geometrical properties of microvessels together with the hydraulic conductivity to serum albumin and the reflection coefficient.
Although the permeability of the venules is expected to be larger than the permeability of the arterioles, in the absence of specific data, and for illustration purposes in the subsequent exercise we assumed the same permeability for both microvessels.

\begin{table}[ht]
\centering
\begin{tabular}{lll}
\\
\hline 
Parameter [unit]                         & Value                   & Reference \\
\hline
$K$ [kg \ sec$^{-3}$ (cm H$_2$O)$^{-1}$] & 2.49 $\cdot$ 10$^{-12}$ & [Michel and Curry, 1999] \\
$\sigma$                                 & 0.85                    & [Michel, 1980] \\
$n$                                      & 0.5                     & [Robinson, 1988] \\
$r_A$ [$\mu$ m]                          & 15                      & [Silverthorn, 2010] \\
$r_V$ [$\mu$ m]                          & 10                      & [Silverthorn, 2010] \\
$\Delta x_A$ [$\mu$ m]                   & 6                       & [Silverthorn, 2010] \\
$\Delta x_V$ [$\mu$ m]                   & 1                       & [Silverthorn, 2010] \\
\hline \\ 
\end{tabular}
\tiny
\caption{Typical values of the parameters used in the computation. $K$ is the hydraulic conductivity for serum albumin, $\sigma$ is the reflection coefficient for serum albumin, $n$ is the porosity, $r$ is the mean radius of the vessel and $\Delta x$ is the vessel thickness. $A$ refers to the arteriolar end of the capillary bed, while $V$ to the venous end.}
\label{tab2}
\end{table}

In addition, venules and arterioles are subjected to different internal hydrostatic pressures and external osmotic pressures. Table \ref{tab1} shows the typical mean pressures in different microvessels.

\begin{table}[ht]
\centering
\begin{tabular}{lccccc}
\\
\hline
Location & $P_c$ & $P_i$ & $\sigma \Pi_c$ & $\sigma \Pi_i$ & $\Delta p$ \\ 
         & [cm H$_2$O] & [cm H$_2$O] & [cm H$_2$O] & [cm H$_2$O] & [cm H$_2$O] \\ 
\hline
arteriolar end & & & & \\
of capillary   & 47.62 & -2.72 & 38.10 & 0.14 & 12.38 \\ 
venular end    & & & & \\
of capillary   & 20.41 & -2.72 & 38.10 & 4.08 & -10.88 \\
\hline \\ 
\end{tabular}
\caption{Mean pressures in human body, taken from [Boron and Boulpaep, 2005]. $P$ represents the hydrostatic pressure, while $\Pi$ is the osmotic pressure and $\sigma$ is the reflection coefficient. 
The subscript $c$ refers to the pressure measured inside the vessel, while the subscript $i$ is measured just outside the vessel.
$\Delta p$ is defined as the difference of the net pressure $p$ between the internal and the external side of the microvessels, i.e. 
$\Delta p = p_c-p_i = (P_c - \sigma \Pi_c) - (P_i - \sigma \Pi_i)$.}
\label{tab1}
\end{table}

The difference of the net pressure $p$ between the internal (subscript $c$) and the external side (subscript $i$) of the microvessels, i.e. 
\begin{equation}
\Delta p = p_c-p_i = (P_c - \sigma \Pi_c) - (P_i - \sigma \Pi_i),
\end{equation}
provides a first rough quantification of the expected flux through the vessel wall per unit area, i.e. the specific discharge. In Table \ref{tab1}, we observe that $\Delta p$ is positive for arterioles (12.38 cm H$_2$O) and negative for venules (-10.88 cm H$_2$O). This leads to a tendency for absorpion at the venular end of the capillary bed, which may be contrasted by the parallel increase of the osmotic pressure within the clefts just downstream the glycocalyx, the membrane coating the internal surface of the endothelial cells \cite{levick:03}. As mentioned before, in the present work we neglect this feedback mechanism. 

\begin{figure}[ht]
\centering
\includegraphics[width=8.5cm]{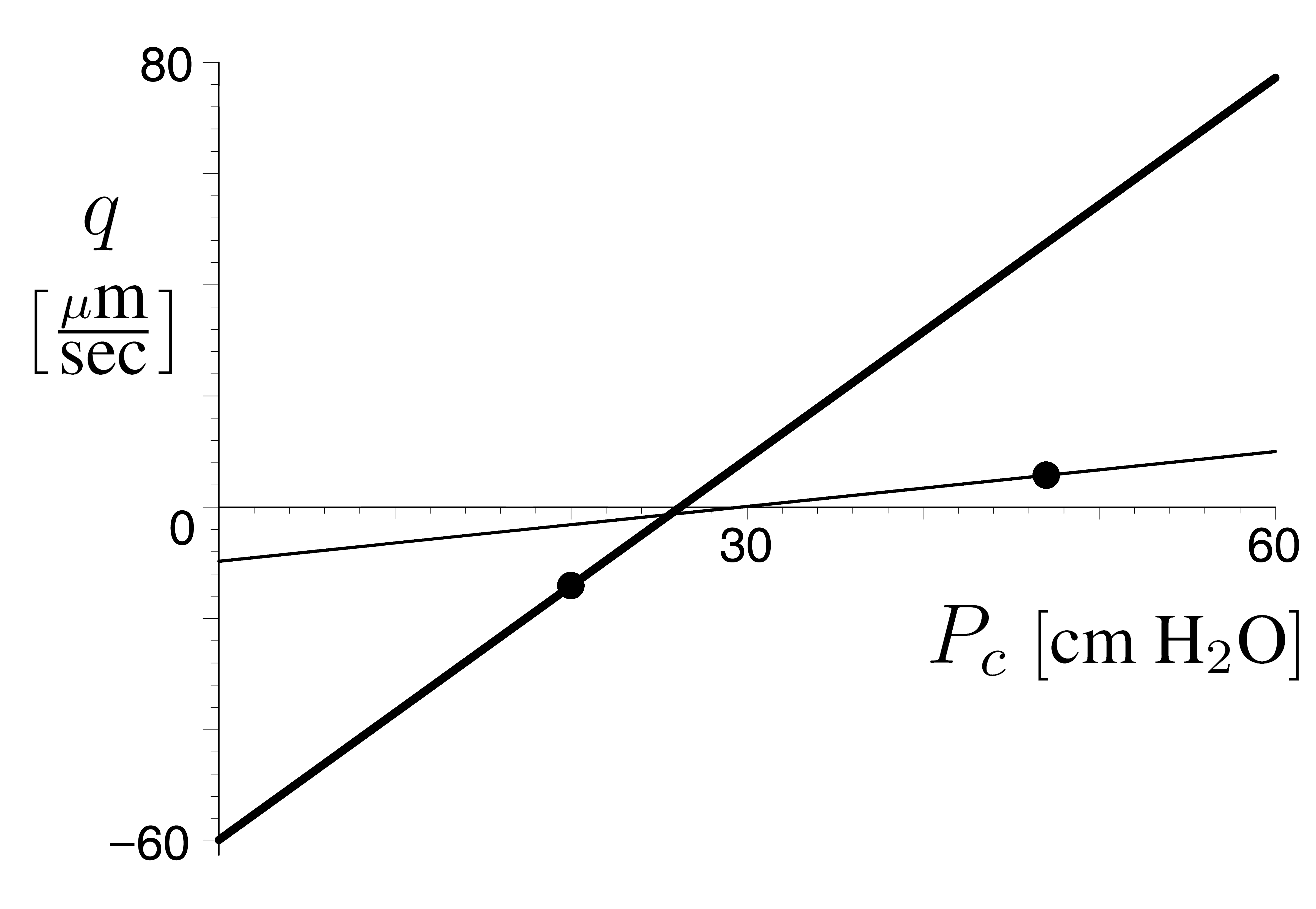}
\smallskip
\caption{Discharge per unit length depending on the internal hydrostatic pressure, in the arteriolar case (thin straight line) and in the venular case (thick straight line). The dots represent the typical values of internal blood pressure in both cases.}
\label{fig1}
\end{figure}

We start by considering the microvessel wall composed by a single layer. 
Figure \ref{fig1} shows the specific discharge $q$ crossing the vessel wall as a function of the hydrostatic pressure $P_c$ for both arterioles and venules. 

In the case in which the vessel wall is composed by only one layer, we can study the behavior of the discharge per unit length and of the travel time of a molecule, assuming that the external pressures $P_i$ and $\Pi_i$ and the internal osmotic pressure $\Pi_c$ are constant.
The internal hydrostatic pressure $P_c$ is the residual pressure, controlled by the cardiac pressure, so we can represent our quantities with respect to it.

For typical values of venular pressure (see the black bullet on the thick straight line in Figure \ref{fig1}), the specific discharge is negative, meaning that venules absorb fluid and the dissolved molecules from the interstitial volume. On the other hand, arteriolar pressure is positive letting the oxygen and the nutrients nourish the surrounding tissues. 

\begin{figure}[ht]
\centering
\includegraphics[width=8.5cm]{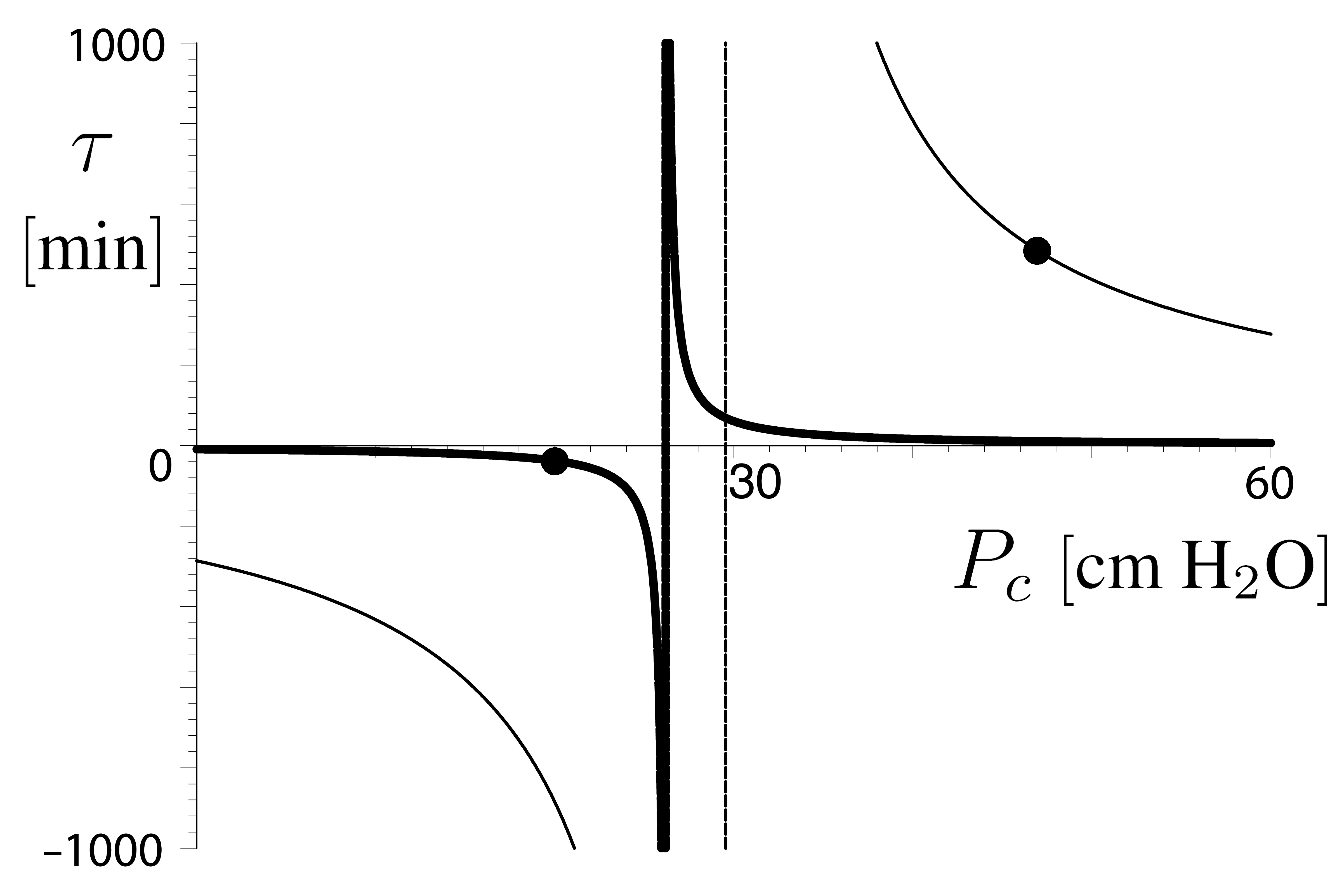}
\smallskip
\caption{Travel time of a molecule of serum albumin, in arterioles (thin curve) and in venules (thick curve). The black bullets represent the
typical values of internal blood pressure in both cases.}
\label{fig2}
\end{figure}

In Figure \ref{fig2}, the travel time $\tau$ of a target molecule (in this case, serum albumin) is depicted with respect to the internal hydrostatic pressure $P_c$ for arteriolar (thin curve) and venular end (thick curve) of the capillary bed. 

For typical values of internal pressure (see the black bullets in Figure \ref{fig2}), $\tau$ is positive for arterioles and negative for venules, reflecting the opposite direction of the flow in the two cases. An increase of the hydrostatic pressure leads to a reduction of $\tau$ for the arterioles. In the case of venules, the same increase leads to a larger travel time $|\tau|$.
Both occurrences may induce a significant alteration of the exchange mechanisms between the interstitial fluid and the cells.  If the hydrostatic pressure increases above a given threshold (about $26$ cm H$_2$O, in the present case) the flux is inverted across the venule wall and the travel time becomes positive, thereby leading to leakage of hematic fluid from the venules into the interstitial volume. Close to this threshold $\tau$ is large, but it reduces rapidly as the hydrostatic pressure increases further. This may provide a plausible explanation for streaks of blood observed in the histology of MS brain plaques \cite{Singh&Zamboni}.  

Finally, we observe that our simple model is in agreement with the early experiments conducted by Landis \cite{Landis} in frog mesenteric
capillaries.

\section{CONCLUSIONS}

We have presented a simplified analytical model of steady-state flow and transport of a target molecule through the wall of microvessels. The advantage of this model is that it allows us to easily explore the explicit influence of the many parameters controlling the process and thereby
avoiding, for the time being, the use of numerical methods.  With this model we have performed a preliminary analysis of the flux across arterioles and venules by using parameters taken from existing studies on mesenteric capillaries. In both cases we computed the time a target molecule (with a given reflection coefficient) spends crossing the  wall, which may provide an indication of the alteration of exchange
mechanisms due to modification of the hydrostatic pressure at the arteriolar  and venular ends.

An increase of the hydrostatic pressure above the value observed in normal conditions leads to an increase of the flux crossing the wall of arterioles and a corresponding reduction of the travel time. In this condition more hydrophilic molecules are released in the
interstitial fluid surrounding the vessel, thereby potentially reducing downstream the availability of such substances needed for the cell metabolism. On the venular side a threshold hydrostatic pressure separates two different ways of functioning. For hydrostatic pressures below such a threshold the flux is negative and the venule absorbs fluid from the interstitial space, while above this threshold the venule leaks hematic fluid to the interstitial space. An increase of the hydrostatic pressure has then a different impact according to the reference hydrostatic pressure. For low reference pressure (i.e. below the threshold) an increase of the hydrostatic pressure leads to a reduction of
the absorption and a parallel increase of the travel time. However, if the reference pressure is larger than this threshold the venule behaves similarly to the arteriole and leaks hematic fluid to the interstitial space with a travel time that reduces rapidly with the increase of the hydrostatic pressure. The impact of these alterations on the cell metabolism may be significant and potentially may be responsible of the suffering status of oligodendrocytes of patients affected by MS and CCSVI.

\providecommand{\bysame}{\leavevmode\hbox to3em{\hrulefill}\thinspace}

\end{document}